\documentclass{llncs}

\usepackage{amsmath}
\usepackage{amssymb}
\usepackage{amsfonts}
\usepackage{theorem}
\usepackage{multicol}

\def\complex{\mathbb{C}}

\def\naturals{\mathbb{N}}

\def\01{\{0,1\}}

\newcommand{\ket}[1]{|#1\rangle}                    % Bra
\newcommand{\proj}[1]{|#1\rangle\langle#1|}         % Projector
\newcommand{\be}{\begin{equation}}
\newcommand{\ee}{\end{equation}}
\newcommand{\bea}{\begin{eqnarray}}
\newcommand{\eea}{\end{eqnarray}}
\newcommand{\bestar}{\begin{equation*}}
\newcommand{\eestar}{\end{equation*}}
\newcommand{\beastar}{\begin{eqnarray*}}
\newcommand{\eeastar}{\end{eqnarray*}}

\newcommand{\spec}{\text{Spec}\;}
\newcommand{\tr}{\text{Tr}\, }
\def\openone{\leavevmode\hbox{\small1\normalsize\kern-.33em1}}

\newcommand{\id}{\openone}    %\text{id}
\newcommand{\U}{{\rm U}}
\newcommand{\GL}{{\rm{GL}}}

\newcommand{\kpsi}{\ket{\psi}}

\newcommand{\kphi}{\ket{\phi}}

\newcommand{\cB}{{\cal B}}

\newcommand{\cH}{{\cal H}}

\newcommand{\cK}{{\cal K}}

\begin{document}
\title{A quantum information-theoretic proof of the relation between Horn's problem and the Littlewood-Richardson coefficients}

\author{Matthias Christandl}

\institute{Centre for Quantum Computation, Department of Applied Mathematics and Theoretical Physics, University of Cambridge, Wilberforce Road, Cambridge CB3~0WA, United Kingdom \newline and  \newline  Arnold Sommerfeld Center for Theoretical Physics, Faculty of Physics, Ludwig-Maximilians-Universit{\"a}t M{\"u}nchen, Theresienstr. 37, 80333 M{\"u}nchen, Germany \newline \email{matthias.christandl@qubit.org }}

\maketitle

\begin{abstract}
Horn's problem asks for the conditions on sets of integers $\mu$,
$\nu$ and $\lambda$ that ensure the existence of Hermitian operators
$A$, $B$ and $A+B$ with spectra $\mu$, $\nu$ and $\lambda$,
respectively. It has been shown that this problem is equivalent to
deciding whether $U_\lambda \subset U_\mu \otimes U_\nu$ for
irreducible representations of $\GL(d, \complex)$ with highest
weights $\mu$, $\nu$ and $\lambda$. In this paper we present a
quantum information-theoretic proof of the relation between the two
problems that is asymptotic in one direction. This result has
previously been obtained by Klyachko using geometric invariant
theory~\cite{Klyachko98}. The work presented in this paper does not,
however, touch upon the non-asymptotic equivalence between the two problems, a result that rests on the recently proven \emph{saturation conjecture} for $\GL(d, \complex)$~\cite{KnuTao99}.

\end{abstract}
\section{Introduction and Results}

 Given three spectra $\mu, \nu$ and $\lambda$, are there Hermitian
operators $A$, $B$ with $$(\spec A, \spec B, \spec A+B)=(\mu, \nu,
\lambda) \ \ ? $$ It is known as Horn's problem to characterise the
set of triples $(\mu, \nu, \lambda)$ which have an affirmative
answer. Those form a convex polytope whose describing inequalities
have been conjectured by Horn in 1962~\cite{Horn62}. In this paper,
we will not be concerned with the characterisation of the polytope
itself which has by now been achieved~\cite{Klyachko98} \cite{KnuTao99}
but rather with the connection of Horn's problem to the
representation theory of $\GL(d, \complex)$. This connection was
first noted by B.~V.~Lidskii~\cite{Lidskii82} and emerges as a
natural twist to Klyachko's work on the inequalities. More
precisely, he proves the following two theorems relating the
admissible spectral triples to the Littlewood-Richardson
coefficients $c_{\mu \nu}^\lambda$. $c_{\mu \nu}^\lambda$ is the
multiplicity of the irreducible representation $U_\lambda$ of
$\GL(d, \complex)$ in the tensor product representation $U_\mu
\otimes U_\nu$ of $\GL(d, \complex)$. $\mu, \nu$ and $\lambda$
denote the highest weights of the respective representations.

\begin{theorem}\label{theorem-converse} If $c_{\mu \nu}^\lambda \neq 0$, then
there exist Hermitian operators $A$ and $B$ such that \beastar
(\spec  A, \spec  B, \spec A+B) =(\mu, \nu, \lambda). \eeastar
\end{theorem}

\begin{theorem} \label{theorem-horn-our}
For Hermitian operators $A$, $B$ and $C:=A+B$ with integral spectra
$\mu$, $\nu$ and $\lambda$, there is an $N \in \naturals$ such that
\bestar c_{N \mu, N \nu}^{N\lambda} \neq 0. \eestar
\end{theorem}

The original proofs of both theorems are based on deep results in
geometric invariant theory. The contributions of this paper are elementary quantum-information-theoretic proofs of Theorem~\ref{theorem-converse} and of the following variant of Theorem~\ref{theorem-horn-our}:

\begin{theorem} \label{theorem-horn-our2}
For all Hermitian operators $A$, $B$ and $C:=A+B$ on
$\complex^d$ with spectra $\mu, \nu$ and $\lambda$, there is a sequence $(\mu^{(j)}, \nu^{(j)}, \lambda^{(j)})$, such
that
\bestar c_{\mu^{(j)}  \nu^{(j)}}^{\lambda^{(j)}} \neq 0 \eestar
and
\beastar
\lim_{j \to \infty} \frac{\mu^{(j)}}{j}&=&\spec A\\
\lim_{j \to \infty} \frac{\nu^{(j)}}{j}&=&\spec B\\
\lim_{j \to \infty} \frac{\lambda^{(j)}}{j}&=&\spec A+B.
\eeastar
\end{theorem}

Theorems~\ref{theorem-horn-our} and \ref{theorem-horn-our2} can be shown to be equivalent with help of the fact that the triples $(\mu, \nu, \lambda)$ with nonvanishing Littlewood-Richardson coefficient form a finitely generated semigroup (see \cite{ChHaMi05} for a similar equivalence in the context of the quantum marginal problem). Here, we choose to prove Theorem~\ref{theorem-horn-our2} since it more naturally fits our quantum information-theoretic approach. The basis of this approach is an estimation theorem for the spectrum of a density operator (Theorem~\ref{theorem-Keyl-Werner})~\cite{KeyWer01PRA}, which has recently been used~\cite{ChHaMi05} \cite{ChrMit05}\cite{Christ05} to prove a relation analogous to the one presented in this paper between the Kronecker coefficient of the symmetric group and the spectra of a bipartite density operator and its margins.

In
1999, Knutson and Tao proved the \emph{saturation
conjecture\index{saturation
conjecture}\label{saturation-conjecture}} for $\GL(d, \complex)$,
i.e.~they proved that
    $$ c_{N\mu, N\nu}^{N\lambda} \neq 0 \mbox{ for some } N \in \naturals \mbox{ implies } c_{\mu
        \nu}^\lambda\neq 0.
    $$
This result implies that the $N$ in Theorem~\ref{theorem-horn-our} can be taken to be one and the equivalence of the two problems is strict and not only asymptotic. The proof appeared in~\cite{KnuTao99} and introduces the
\emph{honeycomb} model. A more compact version of this proof based
on the \emph{hive} model was given by~\cite{Buch00}, and a more
accessible discussion can be found in~\cite{KnuTao01}.

We proceed with the introduction of the necessary group theory and quantum information theory before turning to the proofs.

\section{Preliminaries}

\subsection{Spectrum Estimation}

The tensor space $(\complex^d)^{\otimes k}$ carries the action of the symmetric group $S_k$ which permutes the tensor factors and the diagonal action of $\GL(d, \complex)$: $g\mapsto g^{\otimes k}$. Since those actions commute in a maximal way, the tensor space decomposes in a form known as Schur-Weyl duality:

$$(\complex^d)^{\otimes k} \cong \bigoplus_{\lambda} U_\lambda
\otimes V_\lambda,$$
where $U_\lambda$ and $V_\lambda$ are irreducible representations of $\GL(d, \complex)$ and $S_k$, respectively. The sum extends over all labels $\lambda$ that are partitions of $k$ into $d$ parts, i.e. $\lambda=(\lambda_1, \ldots, \lambda_d)$ where the positive integers $\lambda_i$ obey $\lambda_i\geq \lambda_{i+1}$. As a label of an irreducible representation of $\GL(d, \complex)$, $\lambda$ is a \emph{dominant weight} and as a label of an irreducible representation of $S_k$ it is a \emph{Young frame}.

The following theorem has been discovered by Alicki, Rudnicki and Sadowski in the context of quantum optics~\cite{AlRuSa87} and independently by Keyl and Werner for use in quantum information theory~\cite{KeyWer01PRA}. In~\cite{ChrMit05} a short account of Hayashi and Matsumoto's elegant proof~\cite{HayMat02} of this theorem is given.

\begin{theorem}\label{theorem-Keyl-Werner}
Let $(\complex^d)^{\otimes k} \cong \bigoplus_\lambda U_\lambda
\otimes V_\lambda$ be the decomposition of tensor space according to
Schur-Weyl duality and denote by $P_\lambda$ the projection onto
$U_\lambda \otimes V_\lambda$. Then for any density operator $\rho$
with spectrum $r$ we have \be \label{eq-KeylWerner-1}\tr P_\lambda
\rho^{\otimes k} \leq (k+1)^{d(d-1)/2} \exp \left(-k D
(\bar{\lambda}||r)\right)\ee where $D(\cdot ||\cdot)$ denotes the
Kullback-Leibler distance of two probability distributions and $\bar{\lambda}=(\bar{\lambda}_1, \ldots, \bar{\lambda}_d)=(\frac{\lambda_1}{|\lambda |}, \ldots, \frac{\lambda_d}{|\lambda |})$. $|\lambda |=\sum_i \lambda_i=k$.
\end{theorem}

This theorem can be interpreted as follows: The joint measurement of $k$ copies of the state $\rho$ by projection onto the spaces $U_{\lambda'} \otimes V_{\lambda'}$ will -- with high probability -- result in a measurement outcome $\lambda'=\lambda$ satisfying $\frac{\lambda}{k}\approx r$. $\frac{\lambda}{k}$ is therefore an estimate for the spectrum of $\rho$. Indeed the error exponent in eq.(\ref{eq-KeylWerner-1}) is optimal~\cite{KeyWer01PRA}.

\subsection{Littlewood-Richardson coefficients}

Given two irreducible representations $U_\mu$ and $U_\nu$ of $\GL(d, \complex)$ with highest weights $\mu$ and $\nu$ we decompose the tensor product representation $U_\mu \otimes U_\nu$ of $\GL(d, \complex)$ (here, the group is represented simultaneously with $U_\mu$ and $U_\nu$) into irreducible representations of $\GL(d, \complex)$
\be  \label{LR-def-eq}U_\mu \otimes U_\nu \cong \bigoplus_\lambda c_{\mu \nu}^\lambda U_\lambda,\ee
where $c_{\mu \nu}^\lambda $ denotes the multiplicity of $U_\lambda$ and is known as the \emph{Littlewood - Richardson coefficient}. Since $\GL(d, \complex)$ is the complexification of $U(d)$, the unitary group in $d$ dimensions, we are allowed to -- and will later on -- regard all representations as representations of $U(d)$. The definition of the Littlewood-Richardson coefficient in eq. (\ref{LR-def-eq}) is indeed the standard one. In the proofs below, however, we will work with a different definition given in terms of the
symmetric group: \be
V_\lambda\downarrow^{S_n}_{S_k\times S_{n-k}} \cong \bigoplus_{\mu, \nu} c_{\mu \nu}^\lambda V_\mu \otimes V_\nu.
\ee Here, we restricted the irreducible representation $V_\lambda$ of $S_n$ to the subgroup $S_k \times S_{n-k}$ and decomposed it into products of irreducible representations of $S_k$ and $S_{n-k}$. Observing that $S_n$ is self-dual, i.e. $V_\lambda^\star\cong V_\lambda$, this definition can be put into the following  invariant-theoretic form \be c_{\mu \nu}^\lambda
=\dim(V_\lambda \otimes V_\mu \otimes V_\nu)^{S_k\times S_{n-k}},\ee
where $S_k\times S_{n-k}$ acts simultaneously on $V_\lambda$ and
$V_\mu \otimes V_\nu$. Clearly, this characterisation only applies
to Young frames, i.e. dominant weights with non-negative parts.
The extension to the case of arbitrary dominant weights follows from the
observation that $c_{\mu \nu}^\lambda$ is invariant under the
transformation \be \label{eq-transformation-weights}
\begin{split}   \mu &\mapsto \mu':=\mu+m (1^d) \\
                \nu &\mapsto \nu':=\nu+n (1^d) \\
                \lambda &\mapsto \lambda':=\lambda+(m+n) (1^d) \\
\end{split}
\ee for integers $m$ and $n$, i.e. $c_{\mu, \nu}^\lambda=c_{\mu', \nu'}^{\lambda'}$. $(1^d)$ is short for $(1, \ldots, 1)$ ($d$ ones).

\section{Proofs}

\subsection{From Hermitian Operators to Density Operators}
It will suffice to prove our results for nonnegative Hermitian operators and Young frames, i.e. dominant weights with nonnegative parts. In order to see this, assume that Theorem~\ref{theorem-converse} holds for Young frames and consider an arbitrary triple of dominant weights $(\mu, \nu, \lambda)$ with $c_{\mu, \nu}^\lambda\neq 0$. Choose $m$ and $n$ large enough so that $(\mu', \nu', \lambda')$ defined above has no negative parts. Since $c_{\mu', \nu'}^{\lambda'}\neq 0$ there are positive Hermitian operators $A'$ and $B'$ with $$(\spec  A', \spec  B', \spec A'+B') =(\mu', \nu', \lambda').$$ This equation is equivalent to
$$(\spec  A, \spec  B, \spec A+B) =(\mu, \nu, \lambda),$$
where $A:=A'-m \id$ and $B:=B'-n \id$.
The latter is obtained by subtracting $(m(1^d), n(1^d), (m+n)(1^d) )$ on both sides of the former observing that $\spec (A'-m \id)=\spec A' - m (1^d)$ (similarly for $B$). A similar argument can be carried out for Theorem~\ref{theorem-horn-our2}.

Since we want to use intuition from quantum information theory, we define $p=\tr A / (\tr A+B)$, $\rho^A=A/ \tr A$ and  $\rho^B=B/ \tr B$. The conditions on the spectra of $(A, B, A+B)$ are then equivalent to the conditions on the spectra of $(\rho^A, \rho^B, p \rho^A+(1-p)\rho^B)$, the convex mixture of density operators (i.e. trace one positive Hermitian operators).

We will therefore prove the following two theorems which are equivalent to Theorems~\ref{theorem-converse} and~\ref{theorem-horn-our2} by the above discussion.

\begin{theorem}\label{theorem-converse2}
Let $(\mu, \nu, \lambda)$ be a triple of Young frames with $c_{\mu, \nu}^\lambda \neq 0$. Then there exist quantum states $\rho^A$
and $\rho^B$ such that
\beastar
\spec  \rho^A &=&\bar{\mu}\\
\spec  \rho^B &=& \bar{\nu}\\
\spec \rho^C &=& \bar{\lambda},
\eeastar
where $p=\frac{|\mu|}{|\lambda|}$ and $\rho^C=p\rho^A+(1-p)\rho^B$.
\end{theorem}

\begin{theorem}\label{theorem-horn3}
For all density operators $\rho^A$, $\rho^B$ and $\rho^C=p\rho^A+(1-p)\rho^B$ on
$\complex^d$ with spectra $\mu, \nu$ and $\lambda$ and $p \in
[0,1]$, there is a sequence $(\mu^{(j)}, \nu^{(j)}, \lambda^{(j)})$, such
that
\bestar c_{\mu^{(j)},  \nu^{(j)}}^{\lambda^{(j)}} \neq 0 \eestar
and
\beastar
\lim_{j \to \infty} \bar{\mu}^{(j)}&=&\spec \rho^A\\
\lim_{j \to \infty} \bar{\nu}^{(j)}&=&\spec \rho^B\\
\lim_{j \to \infty} \bar{\lambda}^{(j)}&=&\spec \rho^C.
\eeastar
\end{theorem}

\subsection{Proof of Theorem~\ref{theorem-converse2}}
We assume without loss of generality that $0 < p \leq \frac{1}{2}$. It is well-known that the Littlewood-Richardson coefficients form
a semigroup, i.e. $c_{\mu \nu}^\lambda \neq 0$ and $c_{\mu'
\nu'}^{\lambda'} \neq 0$ implies $c_{\mu+\mu',
\nu+\nu'}^{\lambda+\lambda'} \neq 0$\cite{Elashvili92}\cite{Zelevinsky97}. As a consequence, $c_{\mu \nu}^\lambda \neq 0$ implies $c_{N\mu
N\nu}^{N\lambda} \neq 0$ for all $N$. For every $N$ we will now construct density
operators $\rho^A_{N}$ and $\rho^B_{N}$ whose limits
$\rho^{A}:=\lim_{N\rightarrow \infty} \rho^A_{N}$ and $\rho^{B}:=\lim_{N\rightarrow
\infty} \rho^B_{N}$ satisfy the claim of the theorem.

Fix a natural number $N$, set $n:=N|\lambda|$ as well as $k:=N|\mu|$
and let $p:=\frac{k}{n}$. Since $c_{\mu \nu}^\lambda$ can only be
nonzero if $|\mu|+|\nu|=|\lambda|$, we further have $n-k=N|\nu|$. As
explained above, the nonnegativity of the parts of $\mu, \nu$ and
$\lambda$ allows us to invoke the characterisation of the
Littlewood-Richardson coefficient in terms of the symmetric group:
$$c_{N\mu, N\nu}^{N\lambda} =\dim (V_{N\mu} \otimes V_{N\nu} \otimes
V_{N\lambda})^{S_{k}\times S_{n-k}},$$ where $S_k$ acts on
$V_{N\mu}$, $S_{n-k}$ on $V_{N\nu}$ and $S_k \times S_{n-k}\subset
S_n$ on $V_{N\lambda}$. Now pick a nonzero $\ket{\Psi_N} \in
(V_{N\mu} \otimes V_{N\nu} \otimes V_{N\lambda})^{S_{k}\times
S_{n-k}}$. Consider
\be \label{eq-horn-conv-5}\begin{split} &\cH^{(1)}\otimes \cdots \otimes \cH^{(k)} \otimes
\cH^{(k+1)}\otimes \cdots \otimes \cH^{(n)} \\
&\otimes \cK^{(1)}\otimes \cdots \otimes \cK^{(k)} \otimes
\cK^{(k+1)}\otimes \cdots \otimes \cK^{(n)}, \end{split} \ee
 where $\cH^{(i)}$ and $\cK^{(j)}$ are isomorphic to
 $\complex^d$. Embed the representation $V_{N\mu}$ in $\cH^{(1)}\otimes \cdots \otimes
 \cH^{(k)}$, $V_{N\nu}$ in $\cH^{(k+1)}\otimes \cdots \otimes
 \cH^{(n)}$ and $V_{N\lambda}$ in $\cK^{(1)}\otimes \cdots \otimes
 \cK^{(n)}$. The symmetric group $S_n$ permutes the pairs $\cH^{(i)}\otimes \cK^{(i)}\cong
 \complex^{d^2}$ and its subgroup $S_k\times S_{n-k}$ permutes the first $k$ and the last $n-k$ pairs separately.

Any irreducible representation of the group $S_k\times S_{n-k}$ is
isomorphic to a tensor product of irreducible representations of
$S_k$ and $S_{n-k}$. $\ket{\Psi_N}$ is a trivial representation of
$S_k\times S_{n-k}$ and can therefore only be isomorphic to the
tensor product $V_k\otimes V_{n-k}$ of the trivial representations
$V_k \equiv V_{(k, 0, \ldots, 0)}$ of $S_k$ and $V_{n-k} \equiv
V_{(n-k, 0, \ldots, 0)}$ of $S_{n-k}$. On the first $k$ pairs the
$k$-fold tensor product of $g \in \U(d^2)$ commutes with the action
of $S_k$, and on the remaining pairs it is the $n-k$-fold tensor
product of $g\in \U(d^2)$ which commutes with $S_{n-k}$. Schur-Weyl
duality decomposes the space in~(\ref{eq-horn-conv-5}) into
$$ \bigoplus_{\tau, \tau'} U^{d^2}_\tau \otimes V_\tau \otimes U^{d^2}_{\tau'} \otimes V_{\tau'},$$ so
that
$$ \ket{\Psi_N} \in U^{d^2}_k \otimes V_k \otimes U^{d^2}_{n-k} \otimes V_{n-k},$$
and in terms of projectors onto those spaces
\begin{align*} \proj{\Psi_N} &\leq P_k\otimes P_{n-k} \\
&= [\dim U^{d^2}_k \int_{\complex P^{d^2-1}}\proj{\psi}^{\otimes k} d\psi]\otimes [\dim U^{d^2}_{n-k}\int_{\complex P^{d^2-1}}\proj{\phi}^{\otimes
(n-k)}d\phi].\end{align*}
This directly implies
\beastar
 1&=& \tr \proj{\Psi_N} P_k\otimes P_{n-k} \\
&\leq & \dim U^{d^2}_k \dim U^{d^2}_{n-k}
 \max_{\psi, \phi} \tr \proj{\Psi_N} \proj{\psi}^{\otimes k} \otimes
\proj{\phi}^{\otimes (n-k)}
\eeastar
and therefore guarantees the existence of vectors $\ket{\phi_N}$
and $\ket{\psi_N}$ satisfying
\beastar \label{eq-horn-converse-1}
    \tr [\proj{\Psi_N} \proj{\phi_N}^{\otimes k}\otimes \proj{\psi_N}^{\otimes (n-k)}]
    &\geq &(\dim U^{d^2}_k \dim U^{d^2}_{n-k})^{-1}.
\eeastar Since $\proj{\Psi_N} \leq P_{N\mu} \otimes P_{N\nu} \otimes
P_{N\lambda}$ we have \beastar
    \begin{split} \tr [P_{N \mu} \otimes P_{N \nu} \otimes P_{N \lambda} ]& [\proj{\phi_N}^{\otimes
pn}  \otimes   \proj{\psi_N}^{\otimes (1-p)n} ] \\
        &\geq  \tr \proj{\Psi_N}[\proj{\phi_N}^{\otimes
pn}  \otimes  \proj{\psi_N}^{\otimes (1-p)n}].     \end{split}
\eeastar We define
\begin{align}
\rho^A_N&=\tr_{\cK^{(1)}} \proj{\phi_N}=\tr_{\cH^{(1)}}
\proj{\phi_N} \\
\rho^B_N&=\tr_{\cK^{(k+1)}} \proj{\psi_N}=\tr_{\cH^{(k+1)}}
\proj{\psi_N}
\end{align}
and find, defining $\rho^C_N=p \rho^A_N+(1-p)\rho^B_N$ which
satisfies
$$ \tr P_{N\lambda} (\rho^C_N)^{\otimes
k} \geq \frac{1}{n+1}  \tr P_{N\lambda} (\rho^A_N)^{\otimes pn}
\otimes (\rho^B_N)^{\otimes (1-p)n},
$$
that
\beastar \label{eq-horn-converse-8}
 \tr P_{N\mu}
(\rho^A_N)^{\otimes pn}& \geq &(\dim U^{d^2}_k \dim
U^{d^2}_{n-k})^{-1}\\
    \label{eq-horn-converse-9} \tr P_{N\nu} (\rho^B_N)^{\otimes (1-p)n} &\geq &(\dim U^{d^2}_k \dim
    U^{d^2}_{n-k})^{-1}\\
    \label{eq-horn-converse-10}
    \tr P_{N\lambda}(\rho^C_N)^{\otimes n} &\geq &(n+1)^{-1}(\dim U^{d^2}_k \dim
U^{d^2}_{n-k})^{-1}.
\eeastar
Since $\dim U_n^{d^2}\leq n^{-d^2}$ these are inverse polynomial lower bounds, which, contrasted with the exponential upper bounds from Theorem~\ref{theorem-Keyl-Werner},
$$ \tr P_\mu {\rho^A_N}^{\otimes k} \leq (k+1)^{d(d-1)/2} \exp(-kD(\bar\mu|| r^A))\leq (k+1)^{d(d-1)/2} \exp(-k\epsilon^2/2)$$
and similarly for ${\rho^B_N}$ and ${\rho^C_N}$, imply
$$ || \spec {\rho^A_N} -\bar\mu|| \leq \epsilon$$
$$ || \spec {\rho^B_N} -\bar\nu|| \leq \epsilon$$
$$ || \spec {\rho^C_N} -\bar\lambda|| \leq \epsilon$$
for $\epsilon =O(d\sqrt{(\log N)/N}$). The proof is now completed,
since $N$ was arbitrary and the existence of the limiting
operators is guaranteed by the compactness of the set of density
operators.

\subsection{Proof of Theorem~\ref{theorem-horn3}}
We assume without loss of generality that $0 < p \leq \frac{1}{2}$. If $p$ is rational, consider a positive integer $n$ such that $k=pn$ (otherwise, approximate $p$ by a
sequence of fractions $k/n$). Define purifications
$\kpsi^{AC}$ and $\kphi^{BC}$ of $\rho^A$ and $\rho^B$, respectively such that $p \tr_A\proj{\psi}^{AC}+(1-p)\tr_B \proj{\phi}^{BC}=\rho^C$. Consider the vector
$$ \ket{\tau}=\kpsi^{A_1C_1} \otimes \cdots \otimes \kpsi^{A_{k}C_{k}} \otimes \kphi^{B_1C_{k+1}}\otimes \cdots \otimes \kphi^{B_{n-k}C_{n}},$$
where $\kpsi^{A_jC_j}=\kpsi^{AC}$ and $\kphi^{B_jC_{k+j}}=\kphi^{BC}$. This vector
is invariant under the action of $S_{k}$ permuting systems $A_jC_j$ and $S_{n-k}$ permuting the systems $B_j C_{k+j}$ and is therefore of the form
    $$
        \ket{\tau}=\sum_{\mu, \nu, \lambda}
        \ket{\tau_{\mu \nu \lambda}},
    $$
for vectors $\ket{\tau_{\mu \nu \lambda}} \in U_\mu \otimes U_\nu
\otimes U_\lambda \otimes \big( V_\mu \otimes V_\nu \otimes
V_\lambda\big)^{S_{k}\times S_{n-k}}$.

By Theorem~\ref{theorem-Keyl-Werner}, for all $\epsilon>0$ and
$\mu$ with $\bar\mu \not\in \cB_\epsilon(r^A)=\{ x:=(x_1, \ldots, x_d): ||x -r^A||_1\leq \epsilon \}$ we have
\beastar
    \tr P_\mu (\rho^A)^{\otimes k}
    &\leq& (k+1)^{d(d-1)/2} \exp(-k D(\bar\mu|| r^A))\leq (k+1)^{d(d-1)/2}e^{-\frac{k\epsilon^2}{2 \ln 2}}
\eeastar
where Pinsker's inequality $D(\bar\mu||r^A) \geq
\frac{||\bar\mu-r^A||_1^2}{2 \ln 2} $ was used in the last inequality.
Similar statements hold for $\rho^B$ and $\rho^C$. Together with
\begin{align*}
\tr P_\lambda & \tr_{A_1\cdots A_{k} B_1 \cdots B_{n-k}} \proj{\tau}=\tr P_\lambda  (\tr_A\proj{\psi}^{AC})^{\otimes k} \otimes  (\tr_B\proj{\phi}^{BC})^{\otimes n-k} \\
&=
\tr P_\lambda \frac{1}{n!}\sum_{\pi \in S_n} \pi   (\tr_A\proj{\psi}^{AC})^{\otimes k} \otimes  (\tr_B\proj{\phi}^{BC})^{\otimes n-k} \pi^{-1}\\
&\leq (n+1)\tr P_\lambda(\rho^C)^{\otimes n}
\end{align*}
we obtain (see~\cite{ChrMit05})
\bestar
        \tr P_\mu \otimes  P_\nu \otimes  P_\lambda \proj{\tau}
        \leq (n+1)^{d(d-1)/2} (n+3) e^{-\frac{pn\epsilon^2}{2} }.
\eestar
This estimate can be turned around to give
\be \label{eq-bound-away}
    \sum_{ \substack{(\mu, \nu, \lambda): (\bar\mu, \bar\nu, \bar\lambda) \in \\
    (\cB_\epsilon(r^A), \cB_\epsilon(r^B), \cB_\epsilon(r^C))}} \tr
    P_\mu \otimes P_\nu \otimes P_\lambda \proj{\tau}
    \geq 1- \delta,
\ee
for $\delta :=(n+1)^{d(d+8)/2} (n+3) e^{-\frac{pn\epsilon^2}{2\ln
2} }$, because the number of Young frames with $n$ boxes in $d$
rows is smaller than $(n+1)^d$.

For positive RHS of equation~(\ref{eq-bound-away}) the existence
of a triple $(\mu, \nu, \lambda)$ with $||\bar\mu-r^A||\leq
\epsilon$ (and for $\nu$ and $\lambda$ alike) and $\ket{\tau_{\mu
\nu \lambda}}\neq 0$ is therefore guaranteed. In particular,
$$ c_{\mu \nu}^{\lambda}=\dim (V_\mu \otimes V_\nu \otimes V_\lambda)^{S_{k}\times S_{n-k}} \neq 0$$
holds. The proof of the theorem is completed with the choice of an
increasing sequence of appropriate integers $n$. The speed of
convergence of the resulting sequence of normalised triples to
$(r^A, r^B, r^C)$ can be estimated with $\epsilon=O(d\sqrt{(\log
n)/n})$, a value for which the LHS of eq.(\ref{eq-bound-away}) is
bounded away from zero.

\section*{Acknowledgment}

The technique used in this paper was developed in collaboration with Graeme Mitchison and Aram Harrow in the context of the quantum marginal problem. I would like to thank both of them for many enlightening discussions. The hospitality of the \emph{Accademia di Danimarca} in Rome, where part of this work was carried out, is gratefully acknowledged. This work was supported by the European Commission through the FP6-FET Integrated Project SCALA CT-015714, an EPSRC Postdoctoral Fellowship and a Nevile Research Fellowship of Magdalene College Cambridge.

%\appendix

\bibliographystyle{splncs}
%\bibliography{hornref}

\end{document}